\documentclass[sigconf,screen,natbib]{acmart}

\setcopyright{none}
\acmDOI{10.1145/3540250.3549158}
\acmYear{2022}
\copyrightyear{2022}
\acmISBN{978-1-4503-9413-0/22/11}
\acmConference[ESEC/FSE '22]{Proceedings of the 30th ACM Joint European Software Engineering Conference and Symposium on the Foundations of Software Engineering}{November 14--18, 2022}{Singapore, Singapore}
\acmBooktitle{Proceedings of the 30th ACM Joint European Software Engineering Conference and Symposium on the Foundations of Software Engineering (ESEC/FSE '22), November 14--18, 2022, Singapore, Singapore}

\usepackage{microtype} 
\usepackage[T1]{fontenc}
\usepackage{url}
\usepackage{colortbl}
\usepackage{enumitem}

\usepackage[caption=false]{subfig}
\graphicspath{{figures/}}
\usepackage{graphicx}

\usepackage{listings}

\makeatletter
\lst@Key{countblanklines}{true}[t]%
    {\lstKV@SetIf{#1}\lst@ifcountblanklines}

\lst@AddToHook{OnEmptyLine}{%
    \lst@ifnumberblanklines\else%
       \lst@ifcountblanklines\else%
         \advance\c@lstnumber-\@ne\relax%
       \fi%
    \fi}
\makeatother

\definecolor{codeblack}{rgb}{0,0,0}
\definecolor{lightgray}{gray}{0.98}
\definecolor{codeblue}{rgb}{0.13,0.13,1}
\definecolor{codegrey}{rgb}{0.36,0.35,0.38}
\definecolor{codegreen}{rgb}{0,0.5,0}
\definecolor{codered}{rgb}{0.9,0,0}
\definecolor{purple}{cmyk}{0.41,0.73,0,0.1}

\lstdefinestyle{default}{
    escapechar=\$,
	backgroundcolor=\color{lightgray},
	keywordstyle=[1]\color{codeblue},
	keywordstyle=[2]\color{purple},
	keywordstyle=[3]\color{codered},
 	stringstyle=\color{codegreen},
	commentstyle=\color{codegrey},
	basicstyle=\color{codeblack}\small\ttfamily,
	columns=fullflexible,
    breaklines=true,
	prebreak=\raisebox{0ex}[0ex][0ex]{\ensuremath{\hookleftarrow}},
    xleftmargin=6pt,
    xrightmargin=0pt, 
	frame=none,
    framexleftmargin=0pt,
    framexrightmargin=0pt,
    mathescape=true,
	numbers=left,
	numberblanklines=false,
    countblanklines=false,
	stepnumber=1,
	numbersep=4pt,
	numberstyle=\footnotesize,
	keepspaces=true,
    showspaces=false,
    showstringspaces=false,
	showtabs=false,
	upquote=true,
    tabsize=4
}

\newcommand{\figref}[1]{Figure~\ref{#1}}
\newcommand{\secref}[1]{Section~\ref{#1}}
\newcommand{\tabref}[1]{Table~\ref{#1}}

\usepackage{mdframed}
\usepackage{xpatch}

\makeatletter
\xpatchcmd{\endmdframed}
  {\aftergroup\endmdf@trivlist\color@endgroup}
  {\endmdf@trivlist\color@endgroup\@doendpe}
  {}{}
\makeatother
\newcounter{FindingCounter}
\stepcounter{FindingCounter}
\newcommand{\findings}[1]{
	\begin{mdframed}[backgroundcolor=gray!10,
			linewidth=0.75pt,
			roundcorner=5pt,
			innertopmargin=2mm,
			innerbottommargin=2mm,
			innerrightmargin=2mm,
			innerleftmargin=2mm,
			skipabove=1mm,
			skipbelow=1mm,
			font=\small]
		{\bf{Finding~\arabic{FindingCounter}:}}~ #1
	\end{mdframed}
	\stepcounter{FindingCounter}
}

\newcommand{\para}[1]{\vspace{0.5em}\noindent\textbf{#1}\hspace{1em}}

\begin{document}

\title{An Exploratory Study on the Predominant Programming Paradigms in Python Code}

\author{Robert Dyer}
\email{rdyer@unl.edu}
\orcid{0000-0001-9571-5567}
\affiliation{%
  \institution{University of Nebraska---Lincoln}
  \city{Lincoln}
  \state{NE}
  \country{USA}
}

\author{Jigyasa Chauhan}
\email{jchauhan2@huskers.unl.edu}
\affiliation{%
  \institution{University of Nebraska---Lincoln}
  \city{Lincoln}
  \state{NE}
  \country{USA}
}

\begin{abstract}
Python is a multi-paradigm programming language that fully supports object-oriented (OO) programming. The language allows writing code in a non-procedural imperative manner, using procedures, using classes, or in a functional style. To date, no one has studied what paradigm(s), if any, are predominant in Python code and projects. In this work, we first define a technique to classify Python files into predominant paradigm(s). We then automate our approach and evaluate it against human judgements, showing over 80\% agreement. We then analyze over 100k open-source Python projects, automatically classifying each source file and investigating the paradigm distributions. The results indicate Python developers tend to heavily favor OO features. We also observed a positive correlation between OO and procedural paradigms and the size of the project. And despite few files or projects being predominantly functional, we still found many functional feature uses.
\end{abstract}

\ccsdesc[500]{Software and its engineering~Multiparadigm languages}
\ccsdesc[300]{Software and its engineering~Object oriented languages}
\ccsdesc[300]{Software and its engineering~Functional languages}
\ccsdesc[300]{Software and its engineering~Imperative languages}

\keywords{Python, programming paradigms, empirical study}

\maketitle
\thispagestyle{plain}
\pagestyle{plain}

\section{Introduction}
\label{sec:intro}

Python's popularity has been rising the last few years~\cite{pypl,tiobe}, especially in fields such as Data Science.  The Python language is often taught as the first programming language to students at many universities and has been gaining rapid ground in recent years~\cite{cs1-python}.  The Python language is considered a multi-paradigm programming language, with full support for object-oriented programming.  Every value in the language is actually an object, with many built-in methods (use \lstinline|dir()| to view them, e.g. \lstinline|dir(1)|).  The language also supports several other programming paradigms, to varying degrees, such as: functional programming (lambda, list comprehensions, etc), imperative programming (if, while, etc), procedural programming (def), and aspect-oriented programming (mixins).

Despite the language's popularity, to date no one has investigated if there are \textbf{predominant paradigm(s)} utilized within Python code and projects.  By predominant paradigm, we mean essentially how someone would answer the question ``what paradigm is this code?''  Due to the mixed nature of Python, almost every file is technically ``mixed'' to some degree.  However when humans look at the code, if for example 90\% of the code is utilizing procedures, we would tend to call it ``procedural''.  Thus procedural would be the predominant paradigm for such code.  In cases where people might hesitate to confidently answer such a question, those wind up being labeled ``mixed''.

Questions one might ask about the predominant paradigm of Python code include: Do developers tend to favor object-oriented or functional features when writing code in Python?  Do projects tend to utilize mostly a single paradigm's features or features from many paradigms?  Does the size of the project have any relation to the particular feature(s) it utilizes?

These questions are important to understand as they could help guide future research.  For example, the program comprehension community is interested in how people understand existing Python code~\cite{turner14}, and knowing what paradigm, if any, is predominant could help focus those efforts~\cite{Lucas19} as it may be a confounding factor they could better control for.  Researchers investigating the maintenance of Python code~\cite{Wang2015AnES,Chen20} could benefit knowing the predominant paradigm, as certain techniques may work better (or worse) on specific paradigms~\cite{Lucas19,Wang2015AnES}.  Researchers looking at code smells~\cite{chen16,rahman19} may want to focus on defining new smells in the most predominant paradigm(s) or may want to correlate object-oriented (OO) smells with OO code in Python to see if the frequencies match.  If they don't, figuring out why Python has more/less instances of a smell could lead to better language designs in the future.  Educational researchers could benefit from knowing the predominant paradigm, so when they provide code examples they can focus on language features new programmers are more likely to encounter.

In this work, we investigate the use of language features in the Python language by analyzing over 100k open-source Python projects from GitHub. To do this, we develop an automated classification script using Boa~\cite{boa,boa-website}.  This script is able to take Python source files and classify each as functional, object-oriented, procedural, and/or imperative.  We show that this script agrees with human judgements over 80\% of the time, based on a small sample.  We then leverage this script to investigate how often Python source files (and projects as a whole) utilize each paradigm.

Our results show that overall, Python developers tend to favor an OO paradigm.  However, many single-file projects, the often contain utility scripts, seem to favor a procedural paradigm.  Despite the low number of files and projects classified as functional, we still observe some functional features as highly used as OO features.

We motivate the need to classify a source file's dominant paradigm in the next section.  Then we pose several research questions in \secref{sec:rqs} and outline our approach in \secref{sec:approach}.  We then discuss the results of applying our approach in \secref{sec:eval}.  Then in \secref{sec:threats} we discuss threats to the validity of the study.  In \secref{sec:related} we discuss closely related works.  Finally, we conclude in \secref{sec:conclusion}.

\begin{table*}[ht]
    \centering
    \caption{Classification of Python language features to programming paradigm(s)}
    \label{tab:classification}
\begin{tabular}{c||c|c|c|c}
\textbf{Python Language Feature} & \textbf{Imperative} & \textbf{Procedural} & \textbf{Object-Oriented} & \textbf{Functional\footnote{~}} \\
\hline
\hline
\texttt{if else elif}               & x &   &   &   \\
\texttt{while} loop                 & x &   &   &   \\
\texttt{break}                      & x &   &   &   \\
\texttt{continue}                   & x &   &   &   \\
\texttt{assert}                     & x &   &   &   \\
\texttt{del}                        & x &   &   &   \\
array indexing                      & x &   &   &   \\
\texttt{pass} (inside loop)         & x &   &   &   \\
\texttt{pass} (inside class)        & x &   & x &   \\
\texttt{pass} (inside def)          & x & x &   &   \\
\hline
\texttt{return}			            &   & x &   &   \\
function (\texttt{def})	            &   & x &   &   \\
nested function (\texttt{def})      &   & x &   &   \\
\hline 
\texttt{class} declaration          &   &   & x &   \\
inheritance		                    &   &   & x &   \\
method (\texttt{def})	            &   &   & x &   \\
\texttt{with}		                & x &   & x &   \\
\texttt{try}				        & x &   & x &   \\
\texttt{except}			            & x &   & x &   \\
\texttt{finally}		            & x &   & x &   \\
\texttt{raise}			            & x &   & x &   \\
\hline
\texttt{for} loop                   & x &   &   & x \\
(\texttt{not}) \texttt{in} operator	& x &   &   & x \\
\texttt{yield}				        & x &   &   & x \\
function-as-arg		                &   &   &   & x \\
\texttt{lambda} functions	        &   &   &   & x \\
list comprehension		            &   &   &   & x \\
decorators				            &   &   &   & x \\
generator expressions	            &   &   &   & x \\
iterators (\texttt{\_\_next/iter\_\_()}) & x & x & x & x \\
\texttt{send()} (into generator)	&   & x &   & x \\
\texttt{iter()}			            &   & x &   & x \\
\texttt{map()}		                &   & x &   & x \\
\texttt{sorted()}	            	&   & x &   & x \\
\texttt{filter()}             		&   & x &   & x \\
\texttt{any()}	                 	&   & x &   & x \\
\texttt{all()} 	                    &   & x &   & x \\
\texttt{itertools.*()}	            &   & x &   & x \\
\texttt{functools.*()}	            &   & x &   & x \\
\texttt{enumerate()}		        &   & x &   & x \\
\texttt{zip()}		                &   & x &   & x \\
\end{tabular}
\end{table*}

\begin{figure}[t]
    \centering
\begin{lstlisting}
 class MyNumbers:      # func  oo$\label{ln:name1}$
   x = 1               #       oo        imp
 
   def m(self):        #       oo 
     def m3():         #       oo  proc$\label{ln:proc1}$
       return 1        #       oo  proc$\label{ln:proc2}$
     y = m3()          #       oo  proc$\label{ln:procuse}$
     return y          #       oo
 
   def __iter__(self): # func  oo$\label{ln:name2}$$\label{ln:func1}$
     self.x = 1        #       oo$\label{ln:oo1}$
     return self       #       oo
 
   def __next__(self): # func  oo$\label{ln:func2}$
     y = self.x        #       oo$\label{ln:oo2}$
     self.x += 1       #       oo$\label{ln:oo3}$
     return y          #       oo
 
 x = MyNumbers()       #       oo$\label{ln:instance}$
\end{lstlisting}
    \caption{Python program showing multiple paradigms}
    \label{fig:example}
\end{figure}

\section{Motivation}
\label{sec:motivation}

\newcommand\blfootnote[1]{%
  \begingroup
  \renewcommand\thefootnote{}\footnote{#1}%
  \addtocounter{footnote}{-1}%
  \endgroup
}
\blfootnote{$^1$Functional features as defined in Python's official functional how-to guide~\cite{func-howto}.}
There are many programming paradigms, such as imperative programming which includes object-oriented (OO) and procedural programming, and declarative programming which includes functional, dataflow, reactive, and logic programming.  Most modern programming languages are actually multi-paradigm, in the sense that while they typically have a \textbf{predominant paradigm} they support and emphasize, they also include features from other paradigms.  Python is one example of such a language, as \tabref{tab:classification} shows.

Python is predominantly an OO language, as every value is represented by an object.  For example, if you ask the interpreter \lstinline|type(3)| it tells you the type is \texttt{<class 'int'>}.  But while Python is at its core an OO language, it also supports other paradigms.

This multi-paradigm support often also extends to common Python libraries.  For example, the Django web framework~\cite{django} supports both class-based and function-based (procedural) approaches for implementing views in the framework.  This is partially a result of the evolution of the framework, but both approaches are supported today and there seems to be disagreement on which one people should prefer to use~\cite{django-views,greenfeld2017two}.

This can be confusing to new users of the language, as they are not only learning a new language but might not know how to utilize some of the paradigms.  Classifying such features is also difficult, as some (such as iterators) exhibit properties of more than one paradigm.  In \figref{fig:example} and the discussion below, we show an example and explain how we classify those statements.

\para{Imperative Programming}
In imperative programming, developers specify exactly how something should be computed using statements that modify a program's state.  While procedural and object-oriented programming are both derivatives of imperative, here we classify a statement as imperative only if it does not explicitly use features from another paradigm.  This ignores paradigms assigned to it from its enclosing scope, which we clarify next.

For example, in \figref{fig:example} while the mention of names (e.g. \texttt{MyNumbers} and \texttt{\_\_iter\_\_}on lines \ref{ln:name1} and \ref{ln:name2}) is an imperative feature, because those lines also include class or function declarations we do not mark them imperative.  However, the second line assigning \texttt{x = 1} is imperative, as that line (ignoring its enclosing scope) makes no use of other paradigms.  After marking that statement in isolation, we also look at the statement's enclosing scope and so we \textbf{also} count it as OO, since it is a member of a class.

\para{Procedural Programming}
Procedural programming organizes code into re-usable elements called procedures (note that Python calls these functions -- in this paper, we call them procedures to avoid confusion with functional programming features) that can be called with different arguments.  An example procedure is shown in \figref{fig:example} on lines \ref{ln:proc1}--\ref{ln:proc2}, and it is called on line \ref{ln:procuse}.

Note that in this example, the procedure is nested inside a class method.  Declaring a method and declaring a procedure in Python look very similar, but we consider methods to have a class as its immediate enclosing scope.  This includes class and static methods.

\para{Object-oriented Programming}
Object-oriented programming provides abstractions in the form of classes and instances of those classes called objects.  Users can define classes that contain state (fields) and operations on that state (methods).  Instances can be created and then passed around via reference.  And classes support data hiding in the form of private members.

We consider any declaration of a class, including the entire body of the class, to be a use of OO in Python.  This also includes accessing (lines \ref{ln:oo1}, \ref{ln:oo2}, and \ref{ln:oo3}) and creating objects (line \ref{ln:instance}).  We also classify method calls on objects as OO.  When classifying the paradigm used, one should first look at each statement inside the class body and ignore it exists inside the class.  Does it use functional (lines \ref{ln:func1} and \ref{ln:func2}) or procedural (line \ref{ln:proc1}--\ref{ln:procuse}) features?  If so, it is actually utilizing multiple paradigms and we classify it as such.

\para{Functional Programming}
Python's official documentation includes a guide on writing functionally with the language~\cite{func-howto}.  In this paper, \textbf{when we talk about functional Python programming, this guide is explicitly what we are referring to}.

Functional programming is a declarative style of programming (compared to the prior three, which are all imperative).  This means when writing functional code, developers focus more on describing what they want the program to do and focus less on how they want the program to do it.  The basic module in functional programming is a function.  In Python, functions (in the functional sense) are \texttt{lambda}s.  In functional programming, users create and compose functions together.  Iteration is a core tenant of functional programming and typically accomplished with recursive functions.

It is important to understand how frequently people are using functional features in Python, as it helps inform future changes to the language.  There are several Python Enhancement Proposals (PEPs) accepted or under consideration to add additional functional features, such as the already added generator expression (PEP 289)~\cite{generators-pep} and co-routine support for generators (PEP 342)~\cite{coroutine-pep}, and the proposed structural pattern matching (PEP 634)~\cite{matching-pep}.

\section{Research Questions}
\label{sec:rqs}

In this paper, we focus on the following exploratory questions:

\begin{enumerate}[label={\textbf{RQ\arabic*}},ref={RQ\arabic*}]
    \item \label{rq:stats} \textbf{What is the distribution of programming paradigms for Python projects on GitHub, for each project and for each individual file?} Is one paradigm used more frequently than the others? How often are files and projects utilizing multiple paradigms?

    \item \label{rq:features} \textbf{What are the most and least used features for some programming paradigms?} Are there differences in how language features are used for the most- and least-used paradigms?

    \item \label{rq:size} \textbf{Are project size and predominant paradigm related?} Are size metrics like number of committers, commits, files, or statements related to the predominant paradigm?

    \item \label{rq:evolution} \textbf{How does programming paradigm use change over time?} Do files change their predominant programming paradigm over time?  What are common first/last predominant paradigms?
\end{enumerate}

Next, we outline our approach to answer these questions.

\section{Approach}
\label{sec:approach}

In this section, we outline the approach used to answer the posed research questions.  First, we discuss the tool and dataset used for our analyses.  Next, we discuss a manual classification of Python projects to programming paradigms.  Finally, we discuss an automated solution and compare with the manual approach.

\subsection{Tools and Dataset}

To answer the research questions, we needed a large number of Python projects from many different domains.  We had originally looked at utilizing a Boa~\cite{boa-website} dataset containing Python projects~\cite{sumon2019pydata}, however that data was limited to data-science related projects and we wanted a broader dataset.

As such, we opted to utilize Boa's open-source compiler infrastructure~\cite{boa,boa-compiler} to build our own dataset of Python projects.  We utilized the public GitHub API~\cite{github-api} to query for any project indicating Python as the primary programming language and sorted based on stargazer count.  We then cloned the results and built a dataset.  \tabref{tab:dataset} shows some statistics for that dataset.

\begin{table}[ht]
    \centering
    \caption{Statistics for the Python data analyzed in this paper}
    \label{tab:dataset}
\begin{tabular}{lr}
\toprule
\textbf{Projects}                                           &         101,648 \\
\textbf{All Revisions}                                      &      32,197,017 \\
\textbf{Revisions (with a Python file)}                     &      15,254,331 \\
\textbf{Python Files}                                       &       7,758,882 \\
\rowcolor{gray!15}\qquad\textbf{Python Files (main branch)} &       3,658,391 \\
\textbf{Python File Snapshots}                              &      68,787,597 \\
\textbf{ASTs}                                               & 105,907,774,611 \\
\bottomrule
\end{tabular}
\end{table}

The dataset seemed to contain a large number of files with duplicate ASTs (ignoring comments and whitespace differences), so we identified the duplicates (found by converting the AST into JSON format, then hashing that string similar to \citet{lopes17}) and keep only one instance from each set.  We removed a total of 1,289,982 duplicate files (14.26\% of all files).  All numbers shown in this paper are already deduplicated.

The dataset contains over 100k projects.  There are over 32 million revisions, half of which contain Python files, and 7.5 million unique Python source files comprising over 100 billion abstract syntax tree (AST) nodes.  \tabref{tab:dataset2} shows the per-project statistical distribution for various size metrics, showing the median project has 47 revisions, 52 files with 4,906 statements and 35k AST nodes.

For most analyses, we only example the latest snapshot of the main branch.  For RQ4 and part of RQ3 (that involve committers or commits), we also examine the full history of each file.

\begin{table}[t]
    \centering
    \caption{Per-project statistics for the Python dataset}
    \label{tab:dataset2}
\begin{tabular}{lrrrr}
\toprule
{} &  Revisions &  Python Files &  Statements &        ASTs \\
\midrule
mean &        317 &           627 &     145,678 &   1,041,907 \\
std  &      1,735 &         3,757 &   1,189,335 &   8,689,738 \\
min  &          1 &             1 &           0 &           3 \\
25\%  &         15 &            15 &       1,199 &       8,610 \\
50\%  &         47 &            52 &       4,906 &      35,264 \\
75\%  &        162 &           225 &      26,901 &     191,125 \\
max  &    203,889 &       306,786 &  72,578,872 & 623,549,307 \\
\bottomrule
\end{tabular}
\end{table}

For analysis, we then query this data using the Boa language.  The output of those queries is a specific text format, which we then converted to CSV.  We finally process the CSV files using Python scripts and the Pandas library.  All data, queries, and scripts are provided in our replication package~\cite{replication-package}.

\subsection{Manual Paradigm Classification}
\label{subsec:judgement}

First, we need the ability to classify Python files into the programming paradigm(s) used.  To do this, we took a sample from the previously published Boa Python dataset~\cite{sumon2019pydata} containing a total of 98,537 source files.  From those files (sorted by project URL, then filename), we systematically sampled every $985^{th}$ yielding 102 files.

Some of the projects/files no longer exist on GitHub, so if we found such a case we would simply move to the next file until we found one that still existed.  This resulted in a sample size of 102 source file URLs, giving us a confidence level of 95\% with a confidence interval of +-9.7\%.

We then had three people from our research lab (including the authors) perform a manual classification task.  The task was: given a source file (a URL to view it on GitHub), indicate the predominant programming paradigm(s) you see this file as using.  The raters were explicitly told to ignore imports (we discuss this more in a later section), and were allowed to mark zero, one, or more than one paradigm.  The goal was not to look and see if a single use of a paradigm occurred, but rather to answer the question ``would you call this file functional/OO/procedural/imperative?''  Each human rater was unaware of the judgements from the other two raters at the time of rating.

From this data we computed Fleiss' kappa to measure the inter-rater agreement level.  Fleiss's kappa was used over Cohen's kappa, as there were more than two raters.  The result was a kappa value of 0.717.  This is often interpreted as ``good'' agreement among the raters (with 0.81 being ``very good'').

After all three raters finished their judgements, a discussion round was held.  The discussion focused only on files where there was not unanimous agreement.  Discussion was ordered based on paradigms, meaning all files where the judgement on functional was not 100\% agreement were discussed, then all files for OO, procedural, and finally imperative.  When discussing, the person who had disagreed gave a case for why they rated the file either as or not as that particular paradigm.  Then all three raters were free to update their score.  This repeated for all disagreements.

\begin{table}[ht]
    \centering
    \caption{Human and machine judgements on a sample}
    \label{tab:judgements}
\begin{tabular}{lrr}
\toprule
{} &  Human Judgements &  Machine Judgements \\
\midrule
Imperative &                 5 &                   8 \\
Mixed      &                16 &                  10 \\
OO         &                53 &                  55 \\
Procedural &                28 &                  29 \\
\bottomrule
\end{tabular}
\end{table}

After discussion, we then re-calculated the Fleiss's kappa and found a value of 0.909, meaning ``very good'' inter-rater agreement was reached.  This final set of judgements (shown on the left of \tabref{tab:judgements}) were then used to help calibrate our automated approach, which we describe next.

\subsection{Automatic Paradigm Classification}

Once we had some manual judgements on a sample of data, we were able to more clearly see some of the features that influenced the human judgements.  We first listed as many syntactical (and some API) features from Python that might indicate a particular programming paradigm.  The results of this classification were already shown in \tabref{tab:classification}.  Next, we discussed and assigned each feature in the table to a single, predominant paradigm (our goal was a 1:1 assignment, which was not always possible).

Based on this classification, we wrote a Boa~\cite{boa,boa-website} query to classify each statement into one or more paradigms.  The goal was to map each statement (approximately) based on the classification in the table.  So for example, the following code:

\begin{center}
\begin{tabular}{c}
\begin{lstlisting}[numbers=none,backgroundcolor=\color{white}]
o.setResult(list(map(lambda x: x**3, [1, 2, 3])))
\end{lstlisting}
\end{tabular}
\end{center}

\noindent would be analyzed and generate the following counts:

\begin{center}
\{ 1, 1, 1, 0, 1 \}
\end{center}

\noindent that indicate there was 1 use of functional features (the \texttt{map()} call and the \texttt{lambda}), 1 use of OO features (calling \texttt{setResult} on object \texttt{o}), 1 use of procedural features (the \texttt{list()} call), 0 uses of imperative features, and 1 total statement.  While some numbers might intuitively seem wrong, we explain how we obtain them.

Writing this script required several important decisions.  First, some of the language features being identified are expressions, and we were aiming to classify at the statement level (as an analog to line numbers, which Boa does not contain).  Thus, we had to introduce a function to clamp the counts at the statement level as a single statement might contain 2 (or more) paradigm uses.  This ensured that a paradigm was counted at most once per statement.

Second, Boa simply parses the source code and provides abstract syntax trees (ASTs).  It does not attempt to resolve types or provide type information for analysis.  This complicates several of the mappings.  To get around this issue, we first track all imports in the file to help see what names might be a module.  We also analyze the imports to see specifically if a function was imported from the \texttt{functools} or \texttt{itertools} modules, as we classify all of those functions as functional paradigm.

Third, we had to attempt to assign calls of the form \texttt{expr.m()} as either procedural or OO.  If \texttt{expr} is an object, this is an OO method call on that object.  If \texttt{expr} is a module, it is a normal procedural call.  We utilize the imports to aid this process.

Fourth, we decided to treat imperative a bit special.  We found that almost every line of code, at least in Python, seems to contain at least one imperative feature (such as using a name, assignment, or simple expression arithmetic).  As such, many of the imperative counts were nearly identical to the statement counts.  We decided to make a simple change that stated if the statement directly used another paradigm, do not increment imperative.  Thus the imperative counts we obtain are counts of statements that are fully imperative.

Fifth, we had to decide how to handle large blocks.  For example, when a class is declared how should we count its body?  Clearly the line declaring the class is OO.  We decided the entire body should contribute to OO, regardless of what the individual statements did.  But we did have to handle some corner cases, such as what do we do with a function nested inside of a method?  In that case, it counts as both procedural and OO as it is clearly a procedural method (it does not take self as an argument), but as it is contained within a class's method it is contributing to the functionality of that method and thus is part of the OO nature of the file.

Finally, we had to decide how to handle imports in terms of contributing toward counts.  It was clear these should count as statements toward a file's total statement count.  What was less clear was if it should contribute toward counts for the paradigms.  Part of the issue was that with the lack of resolved types, we are not usually certain what item is being imported.  And in the case of importing a module, should we analyze the whole module to determine what paradigms that used?  What if it was an external library?  These issues led us to decide to \textbf{only} count them as statements, and ignore any potential paradigm count(s).

\subsubsection{Accuracy}

To verify the accuracy of our automated approach, we first built an oracle repository of hand-coded Python files containing various features and named them with the counts we expected.  We then built a local Boa dataset from only that repository and used it as a test-case to identify and fix issues.  We also added regression tests to the repository as we found interesting cases that seemed to give weird results.

We then verified the results in comparison to the human judgements described in the previous section.  Here we used the same sample of 102 files and had the machine classify the file's paradigm.  The goal was to take a file's counts and label it with a single label (either a paradigm, or ``mixed'').  We tried various strategies, all variants of taking the largest number.  Ties or all 0 values result in mixed labels, regardless of strategy.  We set cutoffs, requiring that the winning value must be a certain percentage of the file's statements and tried cutoffs of 0\%, 50\%, 70\%, and 90\%.  We then looked at the performance compared to the human judgements.

Surprisingly, the cutoff at 0\% performed the best - meaning the machine simply looked at the largest number and as long as there were no ties, that was declared the winner.  This agreed with the humans on 82.35\% of the 102 projects.  When we computed the Cohen's kappa to see how well this rating agrees with the human ratings (here it is a single set of values based on majority agreement) we saw a kappa of 0.717, meaning ``good'' agreement.

\subsubsection{Classifying Entire Projects}

The approach described so far only classifies a single file into a paradigm.  We also need the ability to classify entire projects into a paradigm.  For this task, we utilize the counts generated from the prior tool.

First, for each project we add the counts across all their files in the latest snapshot of the main branch.  This gives us a total count across all files, which is what we utilize to classify the project.  Unlike the prior approach however, we do not simply go with the largest value wins.

\begin{figure*}[ht]
    \begin{minipage}[b]{0.99\columnwidth}
    \centering
        \includegraphics[width=\linewidth]{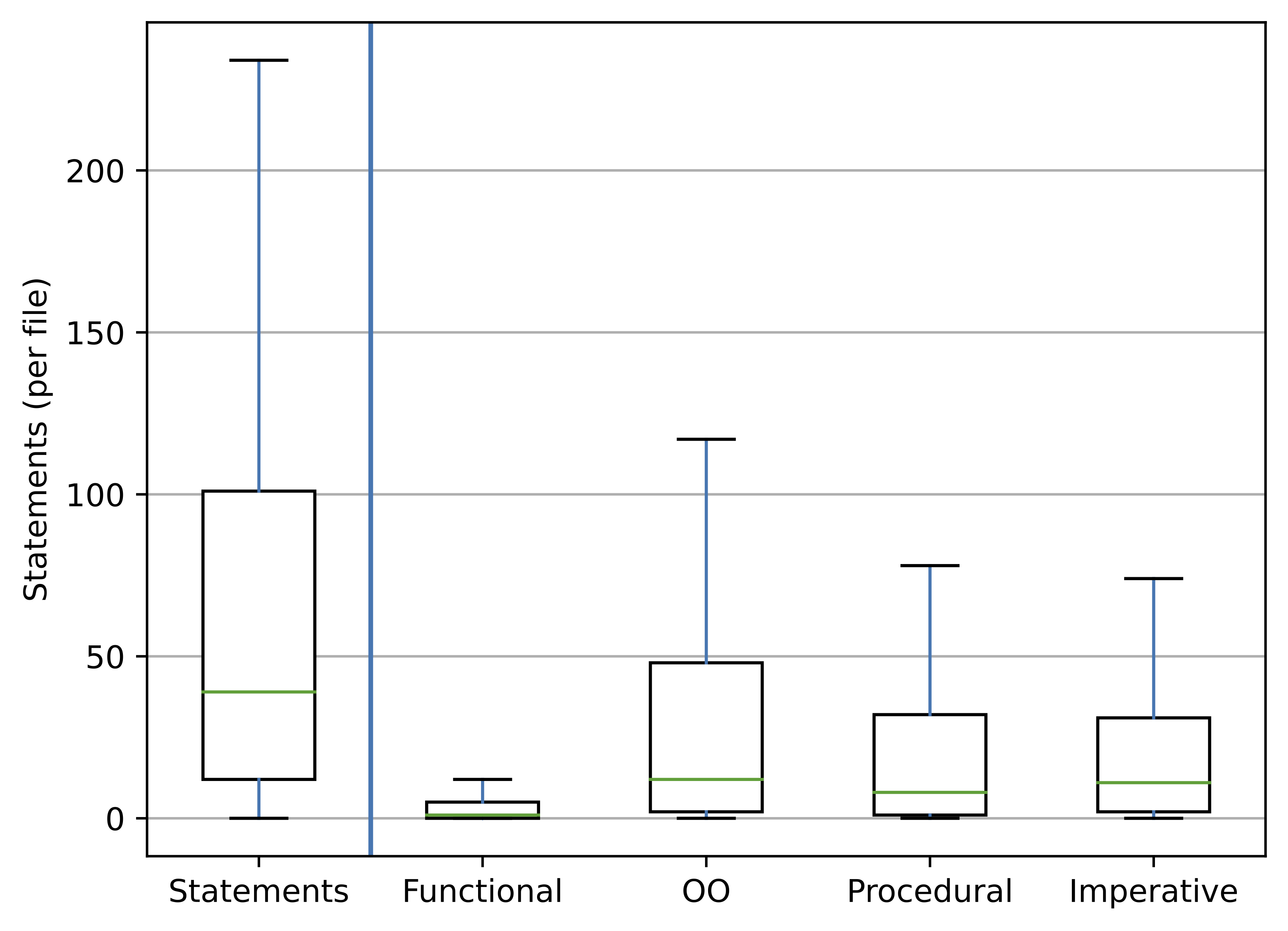}
    \end{minipage}
    \begin{minipage}[b]{\columnwidth}
    \centering
    \small
\begin{tabular}{lrrrrr}
\toprule
{} &  Statements &  Functional &         OO &  Procedural &  Imperative \\
\midrule
mean &      100.10 &        5.58 &      56.31 &       34.81 &       34.13 \\
std  &      969.49 &       38.57 &     583.45 &      152.59 &      762.44 \\
min  &        0.00 &        0.00 &       0.00 &        0.00 &        0.00 \\
25\%  &       12.00 &        0.00 &       2.00 &        1.00 &        2.00 \\
50\%  &       39.00 &        1.00 &      12.00 &        8.00 &       11.00 \\
75\%  &      101.00 &        5.00 &      48.00 &       32.00 &       31.00 \\
max  &  858,430.00 &   38,651.00 & 769,163.00 &   74,194.00 &  858,430.00 \\
\bottomrule
\end{tabular}\\
        Note: the boxplot has outliers removed while the table is a summary of the raw data, including outliers.
        \vspace{3.5em}
    \end{minipage}
    \vspace{-1em}
    \caption{\ref{rq:stats}: Distribution of statements in every unique file (exact duplicates removed) and their assigned classifications}
    \label{fig:rq1-statements}
\end{figure*}

Similar to classifying individual files, if there is a tie for first or if the counts somehow are all 0 we would mark the project as mixed.  For the other cases, we utilized the following algorithm:

\begin{lstlisting}
 if second > 2/3 or largest < 1/3:            # case 1
     return 'mixed'
 if second > 0.5 and largest - second < 0.2:  # case 2
     return 'mixed'
 if largest <= 0.5 and largest - second < 0.1:# case 3
     return 'mixed'
\end{lstlisting}

\noindent While the authors did not perform a full sensitivity analysis to determine the cutoff points, we did try several variations and picked the best performing values based on our intuitions, the data observed, and lengthy discussion.

Note that it is possible for more than 50\% of the statements in a file to be classified to more than one paradigm.  This is due to our classification technique, where we consider each statement and classify what paradigm(s) it uses.  For example, in \figref{fig:example} on line~\ref{ln:proc1} we classify that statement as both OO and procedural.  Depending on the statements, it is possible the entire file is classified as up to 100\% into multiple paradigms.  Thus we needed to consider the edge cases shown in the algorithm above.

Here \texttt{largest} is the largest count (by percentage of total statements) and \texttt{second} the second largest.  When discussing the strategy, the authors quickly realized that some cases were easy to distinguish.  For example, if two (or more) paradigms account for at least 2/3 of the statements then this project is clearly mixing paradigms.  Similarly if the most used paradigm only accounts for 1/3 of the total statements, it was also mixed.  These are the first case.

The second case was if the two highest used paradigms are both over 50\%, and within 20\% of each other then we also considered it mixed.  Finally, the third case was if the most used paradigm is less than half the statements and within 10\% of the second-most used paradigm, we considered it mixed.

\section{Exploratory Study}
\label{sec:eval}

In this section, we discuss the results of analyzing the Python dataset based on the research questions posed.  We also discuss the important findings discovered during analysis and what the implications are of those findings to the community.

\subsection{\ref{rq:stats}: What is the distribution of programming paradigms in Python?}

The first question we want to investigate is simply what the distribution of programming paradigms is for Python files and projects.  To answer this question, we first look at each individual file.  We only consider files from the repository's \textbf{main branch}.  The results are shown in \figref{fig:rq1-statements} as both a box plot and a summary table.

First we look at the statistics on the number of statements in a file.  The median value is 37 statements per file, but there is a lot of variance from 0 (typically empty init files) up to 858k, but most files are relatively smaller.  Next we look at the statistics on the paradigm counts for files.  Here we can see the median values for OO and imperative are the highest, with procedural having a lower median value but otherwise performing similarly to imperative.

\begin{figure}[ht]
    \centering
    \includegraphics[width=\linewidth]{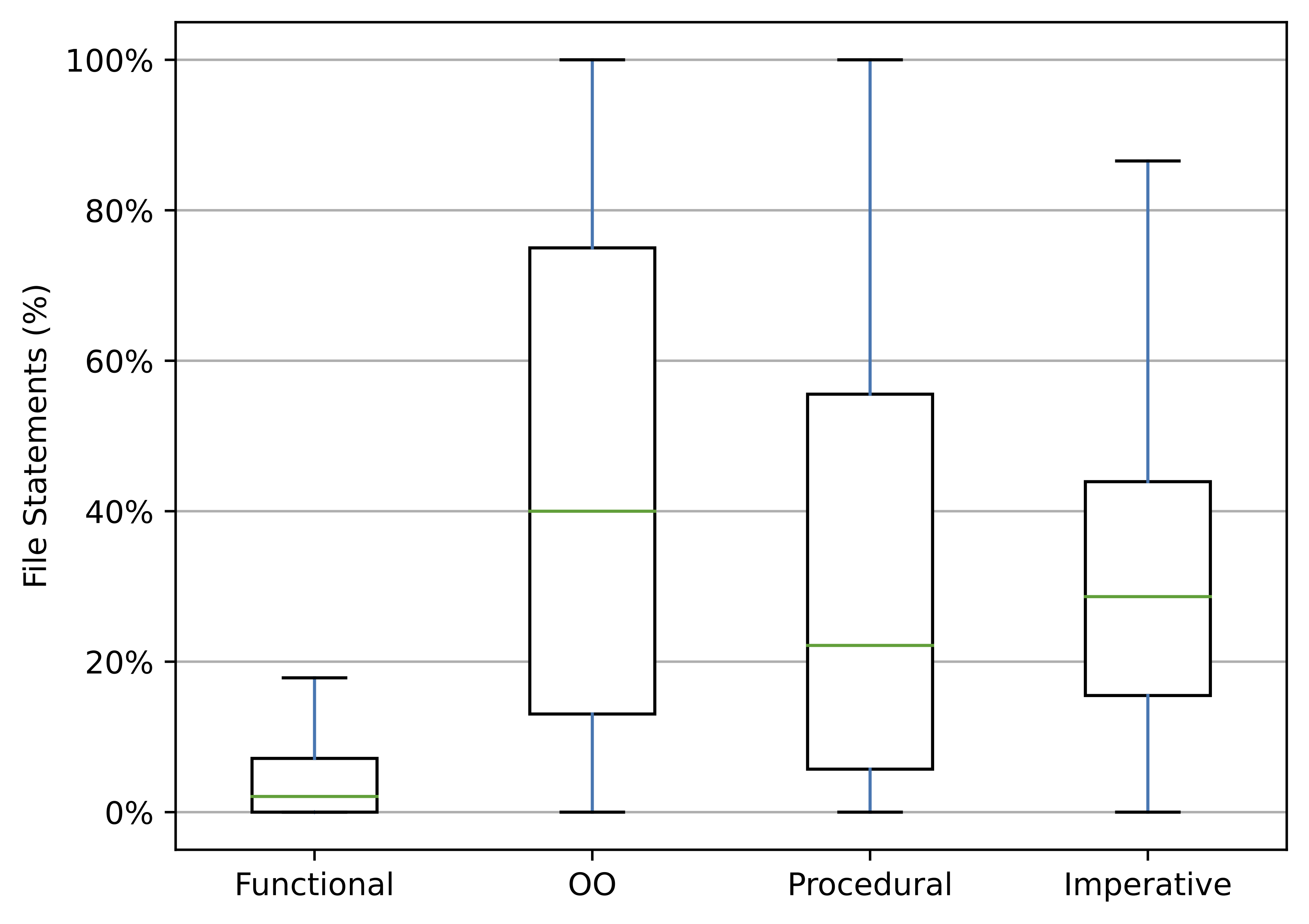}
    \caption{\ref{rq:stats}: percentage of a file that is each paradigm}
    \label{fig:rq1b}
\end{figure}

Next we look at the files in terms of what percent paradigm each file is.  \figref{fig:rq1b} shows the results.  Here it becomes a bit more obvious that, at the file level, OO seems to really dominate the files with the median percentage of a file being around 40\% OO and 30\% imperative.  Functional is the least percentage of files.

\begin{figure}[ht]
    \centering
    \includegraphics[width=\linewidth]{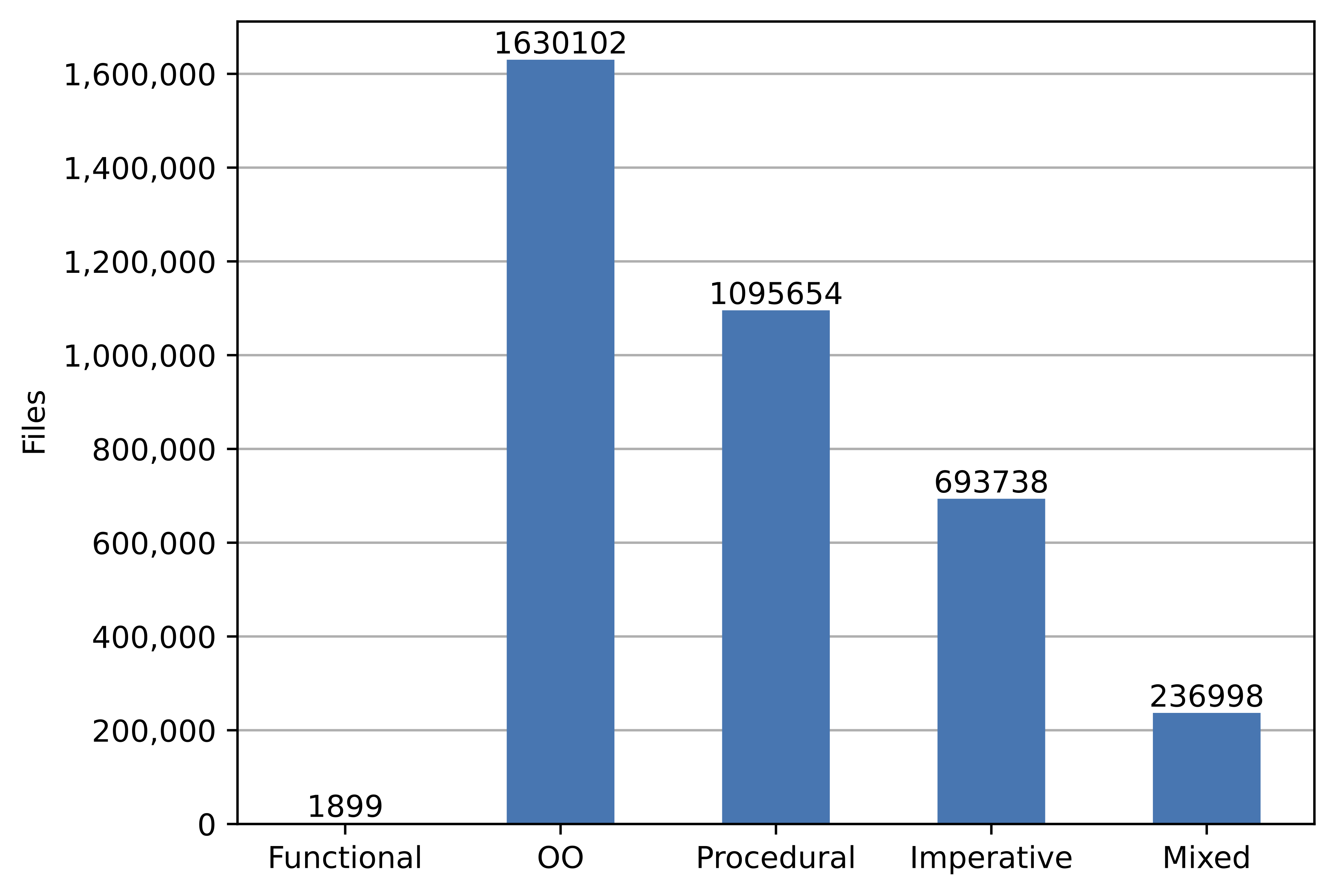}
    \caption{\ref{rq:stats}: Classification of files to paradigm}
    \label{fig:rq1c}
\end{figure}

Finally we use our automated approach to classify each file to a particular paradigm.  The results are shown in \figref{fig:rq1c}.  Here we introduce the new category called ``mixed'', meaning at least two paradigms dominate the file.  Once the script starts classifying files, we can see a clear lead for OO with almost 55\% more files classified OO compared to the second highest (procedural).

\findings{
    Overall, it appears that Python files are predominately written in an OO style.
}

\subsubsection{Classifying Projects}

Next we investigated how projects as a whole classify in terms of paradigm.  \tabref{tab:toys} shows the results.  

As you can see, procedural and OO are the most frequently used paradigms when classifying whole projects.  Given the individual file classifications, this is not too surprising, though imperative being lower than mixed is a bit surprising.  However, this is explained by the strategy used to classify projects being slightly different from files and having more opportunities to mark projects as mixed.

\begin{table}[ht]
    \centering
    \caption{\ref{rq:stats}: Classification of projects}
    \label{tab:toys}
\begin{tabular}{lrrr}
\toprule
{} &  all projects &  no toy projects &  1-file projects \\
classified          &               &                  &                  \\
\midrule
\textbf{Functional} &             1 &                1 &                0 \\
\textbf{OO}         &        45,372 &           39,597 &              497 \\
\textbf{Procedural} &        33,059 &           26,652 &            1,176 \\
\textbf{Imperative} &         5,768 &            4,712 &              310 \\
\textbf{Mixed}      &        17,052 &           14,149 &              387 \\
\bottomrule
\end{tabular}
\end{table}

There is also a possible threat that including ``toy'' projects might bias the results somehow.  When designing the approach, we discussed the idea of filtering out small projects with little history.  We decided to keep them in however, as the questions we are asking are about Python projects as a whole -- including these so-called toys.  That said, we did look into how things might change if we excluded them.  The results are shown in the middle column of \tabref{tab:toys}.  Here we removed any project with a revision count less than 10.

When comparing these results to the full dataset results on the left, we see removing toy projects did not change the trends.  

\findings{
    When considering projects (with or without toys excluded), \textbf{almost no functional projects are found} and \textbf{OO is the most common paradigm} used (over 40\% of all projects) \textbf{followed by procedural} (over 30\% of all projects).
}

We also looked at cases where a project has just a single source file.  The original thought was these might be toys, possibly ``hello world'' style examples where people were simply trying Python out for the first time.  The results are shown in \tabref{tab:toys}.  The results show a mixture of paradigms.  A manual inspection of a small sample of procedural files showed quite a few were utility scripts.

\findings{
    Small (single-file) Python projects are predominately written in a procedural style, and often serve as utility scripts.
}

\subsection{\ref{rq:features}: What are the most and least used features for some programming paradigms?}

When looking at the results of the previous research question, it becomes obvious that some paradigms are highly used in Python (OO, Procedural) and other paradigms are extremely rarely used (Functional), according to the classification of both files and projects.

\begin{figure*}[ht]
    \begin{minipage}[b]{.66\columnwidth}
    \centering
        \includegraphics[width=\linewidth]{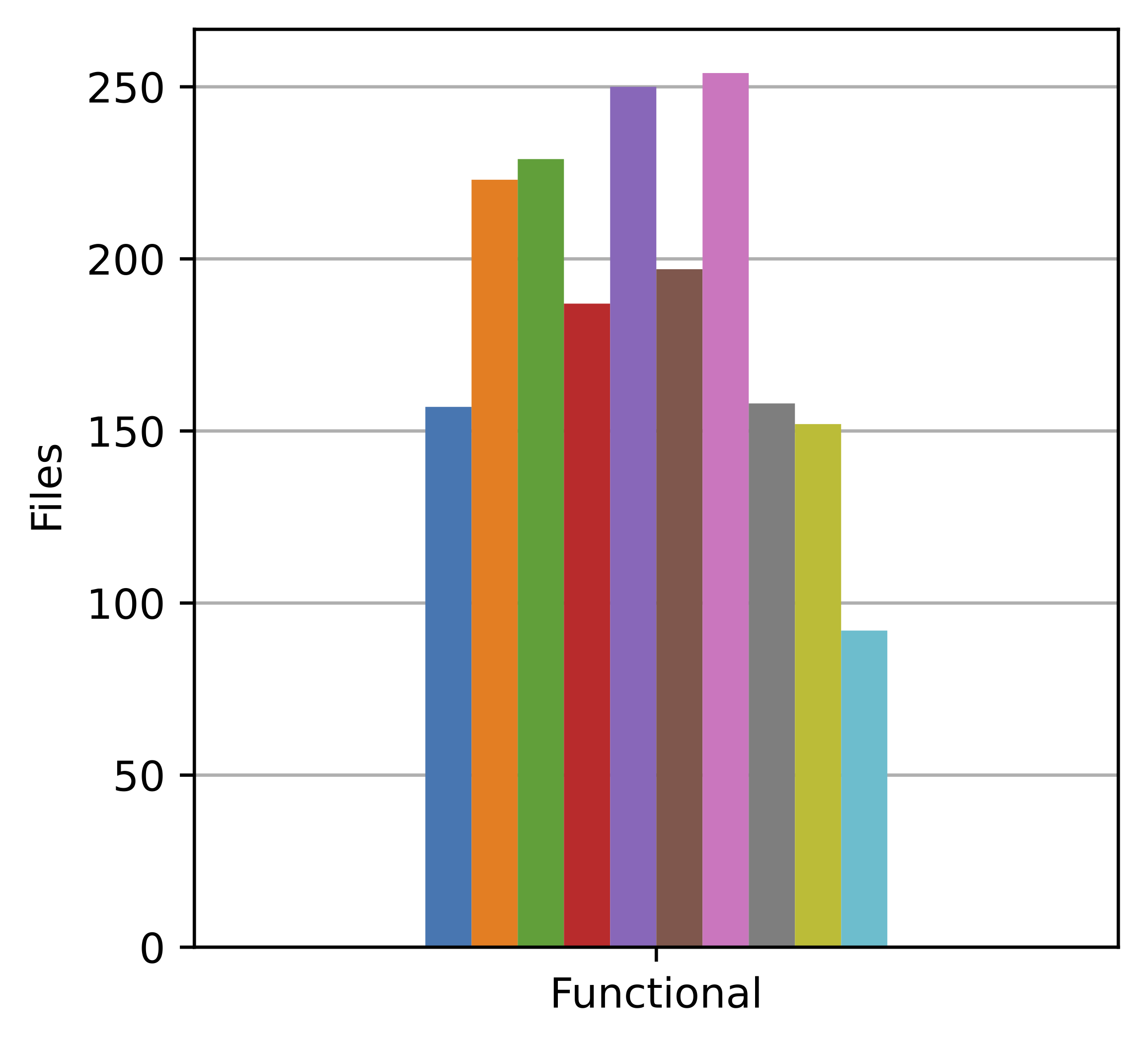}
    \end{minipage}
    \begin{minipage}[b]{1.33\columnwidth}
    \centering
        \includegraphics[width=0.95\linewidth]{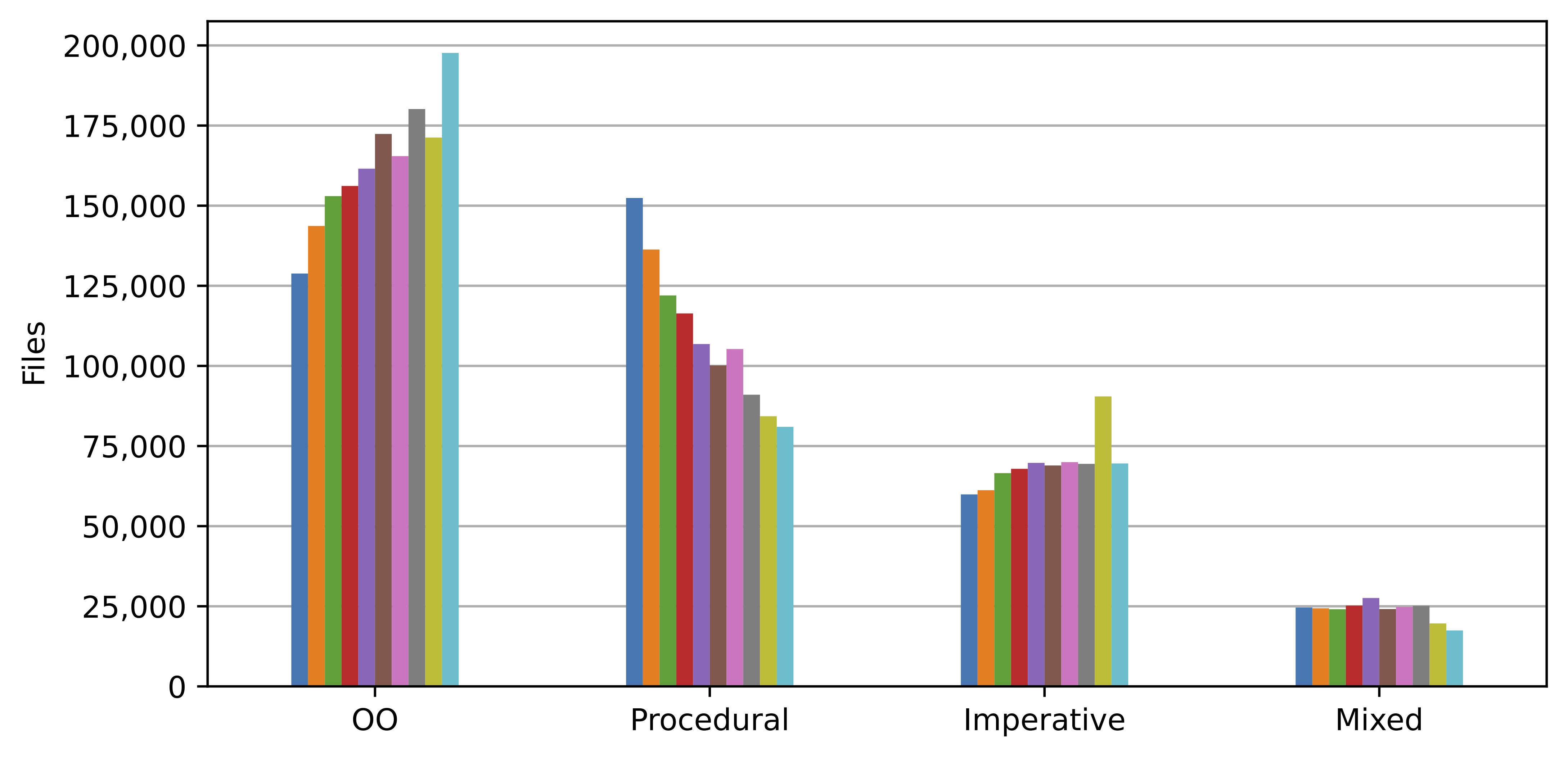}
    \end{minipage}
    \caption{\ref{rq:size}: Is there a relationship between the number of statements in a file and its predominant paradigm?  Here the x-axis is the number of statements in a file, binned into deciles.}
    \label{fig:rq3}
\end{figure*}

Given these results, we were interested in seeing how one of the top paradigms, such as OO\footnote{We chose OO over Procedural as OO has many more features to look at.}, compares to the lowest (Functional) in terms of the use of their individual features.  In other words, given that so many projects and files use OO and so few use Functional, would we also see that functional features are not often used?

We ran a query to find and count how many times the functional and OO features (from \tabref{tab:classification}) were used in the main branch with duplicate files removed.  The results are shown in \tabref{tab:features}.  OO features are highlighted with gray backgrounds.

\begin{table}[ht]
    \centering
    \caption{\ref{rq:features}: Counts for functional and object-oriented programming Python features}
    \label{tab:features}
\begin{tabular}{lr}
\toprule
{} &     count \\
feature                                  &           \\
\midrule
\rowcolor{gray!15} method declarations   & 2,575,397 \\
\rowcolor{gray!15} class declarations    & 1,738,668 \\
\rowcolor{gray!15} class inheritance     & 1,510,161 \\
for-each                                 & 1,405,144 \\
built-in functions (functools/itertools) &   990,333 \\
array comprehensions                     &   729,309 \\
\rowcolor{gray!15} \texttt{try}          &   713,770 \\
\rowcolor{gray!15} \texttt{except}       &   693,096 \\
\texttt{in}                              &   687,249 \\
method decorators                        &   654,500 \\
\rowcolor{gray!15} \texttt{raise}        &   634,313 \\
\rowcolor{gray!15} \texttt{with}         &   559,910 \\
\texttt{not in}                          &   355,880 \\
\texttt{lambda}                          &   300,227 \\
higher-order functions                   &   205,703 \\
\texttt{yield}                           &   128,095 \\
class decorators                         &    90,144 \\
\rowcolor{gray!15} \texttt{finally}      &    83,519 \\
generators                               &    40,931 \\
iterable                                 &     4,323 \\
\bottomrule
\end{tabular}
\end{table}

Given the huge discrepancy between OO and functional in terms of number of files and projects (\figref{fig:rq1c} and \tabref{tab:toys}), and also given the much lower file percentages that are functional compared to OO (\figref{fig:rq1b}) we expected to see substantially more use of OO features compared to functional.  However, the results clearly show that even if files and projects are not being classified as functional, functional code is still frequently used across Python projects.

\findings{
	Despite the low number of files being classified as functional, functional features are still commonly used in Python.
}

\subsection{\ref{rq:size}: Are project size and predominant paradigm related?}

Next we were interested to see if a project's size has an relationship to the programming paradigm it uses.  For this question, we consider different metrics for measuring a projects size, including total number of committers, number of revisions, and total number of statements in all source files.

First, we look at the number of statements.  The results are shown in \figref{fig:rq3}.  We use a histogram to display the values, with data placed into bins at each decile.  Since the maximum values for functional are considerably smaller compared to the other paradigms, we split it out into its own chart on the left.

\begin{figure}[t]
    \centering
    \includegraphics[width=\linewidth]{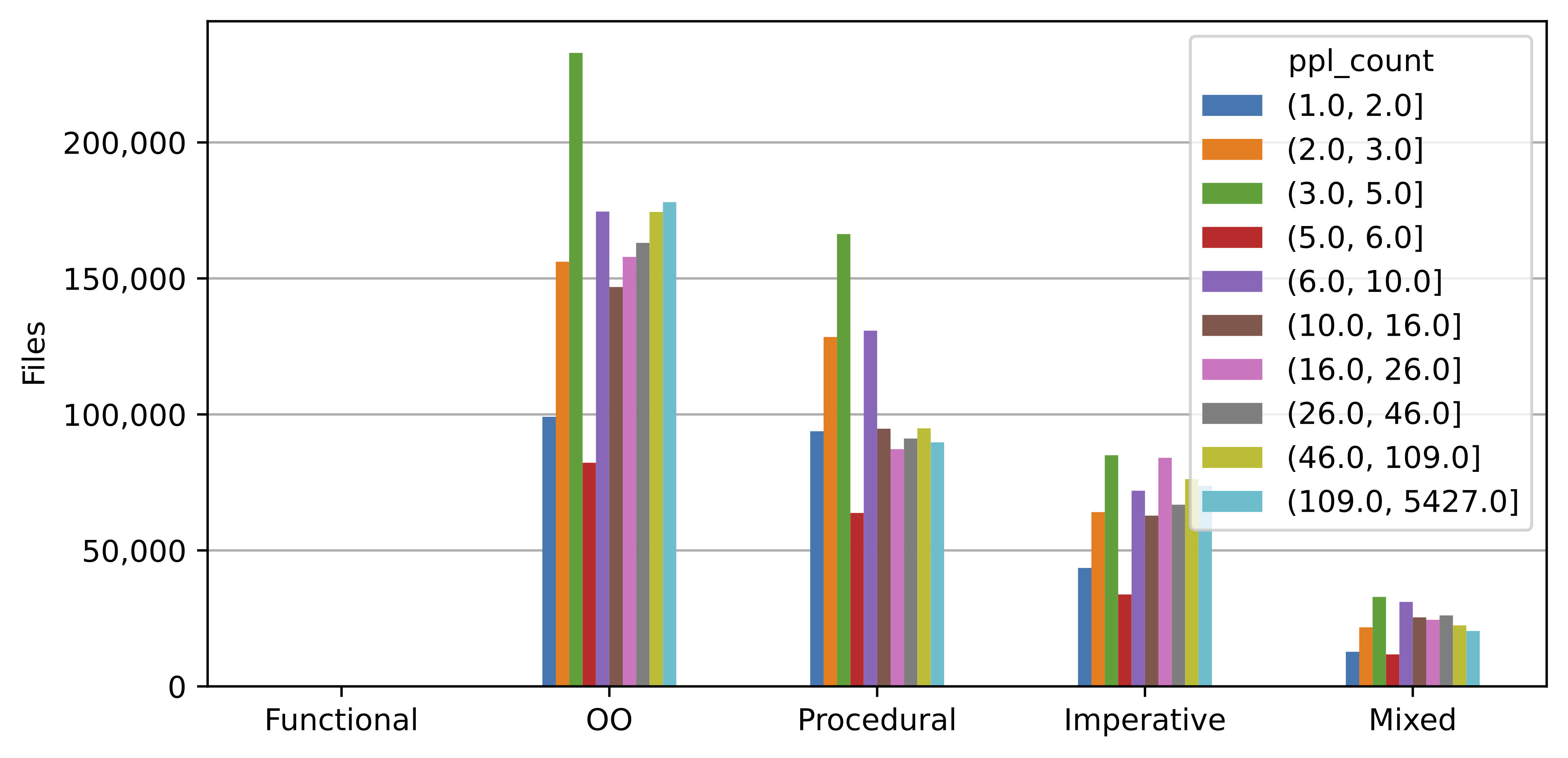}
    \caption{\ref{rq:size}: Number of committers vs paradigm}
    \label{fig:rq3b}
\end{figure}

What we are looking for in these histograms is if the trends seem similar.  In other words, as the project size increases (going right across the x-axis) do we see a similar trend for each paradigm?  What we see are different trends: decreasing, increasing, and no trend.  As the project size is increasing, it appears that the number of files classified as OO or imperative increases showing a positive correlation.  Meanwhile, the number of procedural and mixed files decrease.  And interestingly, the number of functional files seems completely uncorrelated.

\findings{
    The number of statements in a project seems positively correlated to using OO or imperative paradigms, and negatively correlated to using procedural or mixed paradigms.
}

The analysis on the number of committers (\figref{fig:rq3b}) does not show any correlation, most likely due to the extremely skewed data as most projects only have a few committers.  Similarly no correlation was observed when analyzing number of revisions, possibly again due to the skewed nature of the data.  We omit these results from the paper, but they are included in the dataset.

\findings{
    The number of committers and a project's revision history have no correlation to the project's dominant paradigm.
}

\subsection{\ref{rq:evolution}: How do programming paradigm uses change over time?}

The final research question is looking at how the paradigm choice of a file changes over time, if at all.  For this we look at the first and last revision of each unique file \textbf{on all branches} and see if the classification changes or not.  The results are shown in \tabref{tab:rq4evo}.

\begin{table}[ht]
    \centering
    \caption{\ref{rq:evolution}: Do files change their classification from their first revision to their last?}
    \label{tab:rq4evo}
\begin{tabular}{lr}
\toprule
{} &      files \\
changed? &            \\
\midrule
False    & 52,535,151 \\
True     &  8,493,564 \\
\bottomrule
\end{tabular}
\end{table}

As you can see, a small percentage of files (13.8\%) change classification over time.  If a file is using a particular paradigm, the probability of it continuing to use that paradigm is high.  This is good news for developers, as if they are comfortable with the paradigm features used in a file they shouldn't worry about having to learn a new paradigm's features in the future.

We also looked at the 8M files that changed classification to see if there were any trends in how those changes occurred.  The results are shown in \tabref{tab:rq4changed}, with counts indicating how many files were classified as different paradigms in their first and last versions.

\begin{table}[t]
    \centering
    \caption{\ref{rq:evolution}: For files that changed classification: how do they change?}
    \label{tab:rq4changed}
\begin{tabular}{llr}
\toprule
      &            &   files \\
first & last &         \\
\midrule
Functional & OO &   5,540 \\
      & Procedural &   3,991 \\
      & Imperative &   5,305 \\
      & Mixed &   5,931 \\
OO & Functional &   5,677 \\
      & Procedural & 807,967 \\
      & Imperative & 973,349 \\
      & Mixed & 739,926 \\
Procedural & Functional &   3,976 \\
      & OO & 806,868 \\
      & Imperative & 612,296 \\
      & Mixed & 441,578 \\
Imperative & Functional &   5,210 \\
      & OO & 972,512 \\
      & Procedural & 613,026 \\
      & Mixed & 651,935 \\
Mixed & Functional &   5,901 \\
      & OO & 737,745 \\
      & Procedural & 441,416 \\
      & Imperative & 653,415 \\
\bottomrule
\end{tabular}
\end{table}

In \tabref{tab:rq4firstlast} we also show the file counts for a file's first and last classification.  OO was the most common first and last classification, with imperative close behind.

\begin{table}[ht]
    \centering
    \caption{\ref{rq:evolution}: First and last file paradigms}
    \label{tab:rq4firstlast}
\begin{tabular}{lr}
\toprule
{} &     files \\
first      &           \\
\midrule
Functional &    20,767 \\
Imperative & 2,242,683 \\
Mixed      & 1,838,477 \\
OO         & 2,526,919 \\
Procedural & 1,864,718 \\
\bottomrule
\end{tabular}
\begin{tabular}{lr}
\toprule
{} &     files \\
last       &           \\
\midrule
Functional &    20,764 \\
Imperative & 2,244,365 \\
Mixed      & 1,839,370 \\
OO         & 2,522,665 \\
Procedural & 1,866,400 \\
\bottomrule
\end{tabular}
\end{table}

We do not see any obvious trends in the data.

\findings{
    A small number of files (13.8\%) change their classification over time, meaning most files pick a paradigm and stick with it.
}

\section{Threats to Validity}
\label{sec:threats}

Here we consider some potential threats to the validity of our study.

The biggest threat to construct validity is that we rely on inferring paradigm use from already produced code, without speaking with the developers of the project.  So while we may judge a project to be OO based on the analyzed use of various language features, it is possible the developers of the language consider it to be different, e.g. procedural or mixed.  A follow-up study surveying developers based on the results of this study could help confirm or deny that the use of features alone is sufficient to make such a judgement.

The biggest threats to internal validity are the human judgement's (\secref{subsec:judgement}) and the classification of features to paradigm (\tabref{tab:classification}).  Not everyone will agree with how some of the features are mapped.  We attempted to pick an objective mapping, especially for the functional features, by using guidelines from the language designers themselves~\cite{func-howto}.

The human judgement's served as the basis for calibrating the automated approach, so if those judgements were poor the automated approach likely also performs poorly.  While this is a threat, we feel the high inter-rater agreement (0.9) indicates the judgements were high quality and helps mitigate this threat.  However we note the initial agreement was lower (0.7) and after computing the Cohen's kappa to see how well the automated approach agreed with the humans, the human raters started questioning some of their own prior judgements based on the differences from the automated tool.

For example, in one source file\footnote{\url{https://github.com/13o-bbr-bbq/machine_learning_security/blob/08ecb7eea2645a1cecafe21370423fb7393d3bee/Analytics/analyze_kmeans.py\#L18}} the humans judged it as OO and the machine called it mixed. We suspect it was because there was a single statement spanning 9 lines, where each line contained a method call on an object.  Humans probably interpreted this visually large block of text as several OO uses while the machine would count this as a single use.

In another source file\footnote{\url{https://github.com/ARM-DOE/pyart/blob/7584a0b3abe357caa9dcd4b0b50adacffe4af2e0/pyart/aux_io/arm_vpt.py\#L97}}, there were several lines containing code like \lstinline|cfradial._ncvar_to_dict(ncvars['lat'])|. It can be difficult (for both humans and the script!) to determine if this is an OO method call or a procedural call, without global knowledge.  A non-local static analysis would make this result more accurate.

This shows that some projects most likely fall into a gray area, where even humans would most likely disagree on exactly what to classify it as, while the automated tool has to make (sometimes seemingly arbitrary) deterministic choices.  In the future we might utilize machine learning to train a classifier, however currently we lack enough human judgements to build an accurate model.

Another internal threat to validity deals with how the set of features in Python has changed over time.  As such, some features have existed longer than others and this may impact some of the observed results (e.g., the feature counts in \tabref{tab:features}).

A threat to external validity deals with how our results generalize.  While we attempted to utilize a broad range of GitHub projects, we still can't say with certainty those projects are representative of all Python programs/projects and thus the results might not generalize to the population as a whole.  Additionally, the study focused purely on Python, and results almost certainly do not generalize to other programming languages.

\section{Related Works}
\label{sec:related}

In this section we discuss prior works that study the Python language, studies about specific language features, and studies on the use of multiple programming languages.

\subsection{Studies on the Python Language}

\citet{Peng21} analyzed 35 open-source Python projects to see what language features are commonly used.  They found inheritance, decorators, keyword arguments, for loops, and nested classes as the top used language features.  \citet{orru15} looked at the use of inheritance in 51 Python programs.  They observed Python programs have more classes inherited from compared to Java programs.  In contrast to these works, we analyzed almost 100k projects, but with a different goal in mind.  Still, similar to these works, we also found that inheritance, for loops, and decorators are commonly used language features.

\citet{akerblom-dls} looked at the use of polymorphism in 36 Python projects by collecting dynamic traces of the programs.  Their results showed that while many projects use polymorphism, most are actually monomorphic.  \citet{aakerblom2015measuring} looked at 36 Python programs to measure their polymorphic behavior.  They observed many systems, while heavily using polymorphism, are actually monomorphic in behavior.  While these works focuses on the use of inheritance and polymorphism, we look at language features used (including OO inheritance) from a static standpoint and utilize that information for classifying the paradigm(s) used.

\citet{akerblom} used dynamic program tracing to investigate the use of some language features of Python that are difficult to analyze statically, such as dynamically generated code (\texttt{eval}/\texttt{exec}).  While their work looks at specific language features, they are not attempting to categorize programming paradigms those features belong to or determine the predominant paradigm.

\citet{Alexandru18} looked at how developers perceive common idioms in Python, known as the ``Pythonic'' way of solving a problem.  As some idioms might imply a particular paradigm, their work is complementary to our own.  We are not looking at prescribed ways to write code, just observing what was already done.

\citet{Lin16} conducted an empirical study using PyCT focused on 77 kinds of source code changes made by developers. Out of four research questions, one of the questions focused on how often do dynamic features change in source files.  While we do not look at dynamic features, we did investigate how often predominant paradigm changes in source files.

\subsection{Language Feature Studies on Other Languages}

Several other studies have looked at the use of language features (often lambdas) in other programming languages (often Java).  For example, \citet{Mazinanian17} looked at where developers were using lambdas in Java and then surveyed them to find the reasons for using them.  They observed increased use as well as different reasons from increased readability to simulating lazy evaluation.

\citet{Lucas19} looked at if the introduction of lambdas into Java helped with programmer comprehension.  They found contradictory results that lambdas seem to improve program comprehension while simultaneously decreasing readability.

\citet{nielebock19} looked at the use of lambda expressions to aid writing concurrent object-oriented code in Java, C++, and C\# and found that in general, programmers appear to not favor the use of lambdas in the context of concurrent code.

\citet{petrulio21} looked at the support for lambdas in the Java ecosystem and found many top APIs do not yet support functional interfaces.
\citet{Zheng21} looked at why developers remove existing lambdas in Java.  They found seven common reasons to remove lambdas, including reasons such as readability and performance.  They also recommended places to avoid introducing lambdas.

These works focused on a single language feature, while our study focused on identifying the predominant paradigm.  Our work could help focus future studies by indicating which paradigm(s) are most common and thus which feature(s) might be best to study.

\subsection{Multi-language Studies}

\citet{Kochhar16} studied the relationships between multi-language usage and bug categories. They observed if the use of multiple languages causes more bugs. They built regression models to study the correlation of using different languages on the number of bug fixing commits. They noted that languages, when used with other languages, can make software more bug prone.

\citet{Mayer15} investigated the use of multiple programming languages in 1k open-source projects. They used association-rule mining to infer how project size and number of commits related to the languages used. They also discovered several groupings of language ecosystems.

\citet{uesbeck_et_al} addressed the issues programmers face while using multiple programming languages. They performed a control study to determine what kinds of problems programmers face when switching between multiple programming languages, such as Java and SQL.  The study was small and designed as a pilot for future studies.

\citet{Bunkerd19} investigated 50 Python projects to see if the history of developers on those projects and their experience with other programming languages affects the naturalness of their Python code.  They show that greater diversity of contributor programming experience can impact and make the code less natural.

\citet{CHAKRABORTY21} studied Q\&A on Stack Overflow for three languages: Go, Swift and Rust. They monitored the challenges developers initially faced while working with new languages, in comparison to more mature languages. They conducted the study by extracting features, understanding the developer's background as a contributor. The study promoted better design for languages and documentation by the sponsors/owners.

These works study the use of multiple languages, while we focus on the use of multiple paradigms within a single language.

\section{Conclusion}
\label{sec:conclusion}

Python is a multi-paradigm programming language, but to date no one has investigated what paradigms are predominant in Python code.  In this work, we saw that many files and projects favor the OO paradigm.  Single-file projects appear to be utility scripts and favor a procedural paradigm.  The size of the project in terms of statements is positively related to using OO, and negatively related to using functional, procedural, or mixed paradigms.  Finally, we saw the vast majority of files rarely change predominant paradigm over time, providing stability to developers working on those files.

In the future we hope to have a follow-up study to investigate in more detail exactly how humans classify the predominant paradigm of a file.  While it was easy for humans to classify results that are polarized, the middle area was much more difficult.  We would like to perform a survey to investigate why that was the case.

\section*{Acknowledgments}
We thank Samuel W. Flint for helping brainstorm ideas and labeling some of the data.

\section*{Data Availability}
The Boa queries and their outputs, human judgements, and processing scripts are available in a replication package~\cite{replication-package} on Zenodo.

\balance
\bibliographystyle{ACM-Reference-Format}
\bibliography{refs}

\end{document}